\documentclass[aps,showpacs,preprint,superscriptaddress]{revtex4}
\usepackage{graphicx}
\usepackage{subfigure}
\usepackage{float}
\usepackage{amsmath}
\usepackage{amsfonts}
\usepackage{color}
\usepackage{bm}
\usepackage{bbm}
\usepackage{txfonts}
\usepackage{array}
\usepackage{amssymb}

\begin{document}

\title{Schwinger pair production in inhomogeneous electric fields with symmetrical frequency chirp}
\author{Melike Mohamedsedik}
\affiliation{Key Laboratory of Beam Technology of the Ministry of Education, and College of Nuclear Science and Technology, Beijing Normal University, Beijing 100875, China}
\author{Lie-Juan Li}
\affiliation{Key Laboratory of Beam Technology of the Ministry of Education, and College of Nuclear Science and Technology, Beijing Normal University, Beijing 100875, China}
\author{B. S. Xie \footnote{bsxie@bnu.edu.cn}}
\affiliation{Key Laboratory of Beam Technology of the Ministry of Education, and College of Nuclear Science and Technology, Beijing Normal University, Beijing 100875, China}
\affiliation{Beijing Radiation Center, Beijing 100875, China}
\date{\today}

\begin{abstract}
Pair production in inhomogeneous electric fields with symmetrical frequency chirp is studied numerically using the Dirac-Heisenberg-Wigner formalism. We investigate high- and low-frequency modes and consider two carrier envelope phases.
Momentum spectrum is sensitive to chirp causing different interference effect for different spatial scales as well as the carrier phase of the external field. The reduced particle number is in general enhanced with increasing chirp.
The effect of spatial scale of the field on the reduced particle number is also examined. It is found that it is enhanced at small spatial scale but is almost unchangeable at large spatial scales for the considered field parameters. On the other hand, at small spatial scale, the reduced particle number is enhanced by one or two orders when chirp is applied with the exception of cosine low-frequency field which is only a few times larger.
Moreover it is found that the reduced particle number is further increased by symmetrical chirp at about two times by comparing to the usual asymmetrical chirp in high frequency field.
\end{abstract}
\maketitle

\section{Introduction}

Electron-positron pair creation from vacuum in intense electromagnetic fields, known as the Schwinger effect, is a nonperturbative phenomena in quantum electrodynamics (QED) challenging both theoretically and experimentally \cite{Sauter:1931zz,Heisenberg:1935qt,Schwinger:1951nm,Nikishov,breyzin,Marinov,Ritus,Nikishov1985}.
As is identified by Sauter and later formulated by Schwinger within QED, the quantum vacuum decays into real particles through tunneling when the external field strength reaches the critical value $E_{\rm{cr}} =  {m_e^2c^3} / {e\hbar} \approx 1.3 \times 10^{18}  {\rm V}/{\rm m}$ \cite{Schwinger:1951nm}.
However, the corresponding critical field intensity $I_{\rm{cr}}=4.3 \times 10^{29}{\rm W}/{\rm cm^{2}}$ is yet demanding for the current laser technology and the Schwinger effect still awaits an experimental verification. Pair production of vacuum have been investigated for many simple but important fields. For example, the constant external field including the magnetic field is reexamined in \cite{Nikishov} and an alternating electric field has been also researched in \cite{Marinov,breyzin}. In particular, Ritus and Nikishov gave the presentation analytically on the pair production problem in the frame of QED by using the scattering matrix approach \cite{Ritus,Nikishov1985}. Recent theoretical investigations indicate that vacuum pair production can be achieved through various enhancement mechanisms via careful shaping of the laser pulses with intensities one or two orders below the critical value \cite{Schutzhold:2008pz,Bell:2008zzb,DiPiazza:2009py,Bulanov:2010ei}, which brings hope for the detection of the vacuum pair creation with the advanced laser facilities in the near future \cite{Ringwald:2001ib,Heinzl:2008an,Marklund:2008gj,Pike:2014wha}.

Pair creation process, especially the particle momentum distribution, is very sensitive to external field parameters because of its nonlinear nature \cite{Hebenstreit:2009km,Dumlu:2010vv}.The momentum spectrum of the created pairs may provide detail about pair creation dynamics as well as the external field shape\cite{Akkermans:2011yn,Dumlu:2010ua,Dumlu:2011rr}.Many research focus on temporal electric fields through changing of one or more field parameters\cite{Xie:2017,Li:2015cea,Li:2017qwd,Hebenstreit:2009km,Dumlu:2010vv,Abdukerim:2013vsa,Olugh:2018seh}, such as field  frequency \cite{Olugh:2018seh}, carrier phase \cite{Abdukerim:2013vsa}, and analyze effects of these arguments on the momentum spectrum of created particles.

Moreover, it is also well known that the chirp plays an important role in strong field QED. For example, high intensity laser pulse itself, which we need it for the study of pair production, are obtained by chirped pulse amplification (CPA) technique \cite{Strick} in laboratories. Chirp effects on pair production have been widely studied in the homogeneous fields \cite{Dumlu:2010vv,Olugh:2018seh}. To our knowledge, current research has been limited on the asymmetrical chirp effects in the spatially inhomogeneous field set up \cite{Ababekri:2020}. Recently, we have studied symmetrical chirp effects in homogeneous electric field \cite{Wang}.

On the other hand, interest for the effect of spatial inhomogeneity on the vacuum pair production is triggered by the advances in theoretical works and the numerical computation techniques \cite{Hebenstreit:2011wk,Kohlfurst:2017hbd,Ababekri:2019dkl,Ababekri:2020,Kohlfurst:2017git}. These works have provided new features of finite spatial field pair production, such as self-bunching effect \cite{Hebenstreit:2011wk}, ponderomotive force \cite{Kohlfurst:2017hbd}, spin field interaction \cite{Kohlfurst:2017git} and so on. Recently, Kohlf\"{u}rst has studied vacuum pair production in strong electromagnetic field, by considering both of electric and magnetic field with spatial inhomogeneity and has revealed the competitive results between the multiphoton and tunneling process \cite{Kohlfurst:2020,Kohlfurst:2017git}. All these findings emphasize the importance of considering spatial variations for vacuum pair production in the strong external field.

In this paper, we study the vacuum pair creation in symmetrical chirped fields at finite spatial scales. Our approach is based on the Dirac-Heisenberg-Wigner(DHW) formalism, which is first introduced to study the vacuum decay to pair production   \cite{Bialynicki-Birula}. This approach has been a powerful tool to study the pair production \cite{Hebenstreit:2011wk,Kohlfurst:2017hbd,Ababekri:2019dkl,Ababekri:2020,Kohlfurst:2017git,Li:2015cea,Xie:2017}. We consider oscillating field forms in cosine and sinusoidal form. We find that momentum spectrum is very sensitive to symmetrical chirp which exhibits different interference effect for different spatial scales as well as the carrier phase of field. The reduced particle number is in general enhanced with increasing chirp. In particular, at small spatial scale, the reduced particle number is enhanced at most by one or two orders when chirp is applied. Moreover it is found that the reduced particle number is further increased by symmetrical chirp at about two times compared to asymmetrical chirp.
We use natural units ($\hbar=c=1$) and express all quantities in terms of the electron mass $m$.

\section{Model of Background Field}\label{fields}

We consider the following spatially inhomogeneous 1+1 dimensional electric field model
\begin{equation}\label{FieldMode}
\begin{aligned}
E\left(x,t\right)
&=E_{0} f \left( x \right ) g\left( t \right )\\
&=\epsilon \, E_{cr} \exp \left(-\frac{x^{2}}{2 \lambda^{2}} \right ) \exp \left(-\frac{t^{2}}{2 \tau^{2}} \right ) \cos(b |t| t + \omega t + \varphi),
\end{aligned}
\end{equation}
where $\epsilon E_{cr}$ is the field strength, $\omega$ is the original center frequency, $\lambda$ and $\tau$ are the spatial and temporal scales, $b$ is chirp parameter which introduces a linear variation in the frequency as time changes and $\varphi$ is the carrier phase. 
For high frequency field, we choose frequency $\omega=0.7m$, temporal pulse length $\tau=45m^{-1}$, while we choose $\omega=0.1m$, $\tau=25m^{-1}$ for low frequency field. For nonzero chirp parameter $b$, it results in time dependent effective frequency $\omega_{\text{eff}}(t)=\omega+b|t|$ that it is always larger than field frequency in the relevant time interval $-\tau \le t \le \tau$. In this paper, the values of chirp $b$ are expressed as the form $b=\alpha\omega/\tau$ ($\alpha \ge 0$), we set the theoretical upper limit for the chirp parameter by letting effective frequency to be around the critical frequency $\omega_{\text{eff}}(\tau) \sim 1.0m$ for high frequency field. In the low frequency field, the maximum chirp value is determined by the condition $\omega_{\text{eff}}(\tau)\tau \sim \mathcal{O}(1)$. These conditions have been discussed in previous works \cite{Ababekri:2020,Kohlfurst:2017git}.

This idealized standing wave model Eq.\eqref{FieldMode} can be regarded to be constructed by two counter propagating laser pulses with the needed symmetrical frequency chirping so that the effect of the magnetic field is automatically vanished. By the way, because of the frequency is of the order of electron mass, which lies in the X-ray regime and so cannot be realised by current lasers, while we hope that it will be realized in future as the rapid development of laser technology. Since the direction of the electric field is along the $x$- axis and the particle is created in the field direction, the study can be simplified as $p_{\perp}=0$.

\section{The DHW formalism}\label{method}

The DHW formalism is one of the quantum kinetic approaches that is widely adopted to study vacuum pair production in multidimensional electromagnetic fields \cite{Kohlfurst:2015zxi,Hebenstreit:2011pm}. It enables one to obtain phase space information about the particles created in vacuum, so that much can be learned about Schwinger pair production process. However, the deficiency of this method is the complicated computational process because one has to solve a set of multidimensional partial differential equations. Therefore, we choose to rely on numerical simulation method used in the previous work \cite{Kohlfurst:2015zxi}.

We focus on pair production in time dependent inhomogeneous electric fields given in Eq. \eqref{FieldMode} such that electric fields are inhomogeneous on $x$- axis, and the complete DHW equations of motion are reduced to the following $4$ differential equations \cite{Kohlfurst:2015zxi}
\begin{align}
 &D_t \mathbbm{s} - 2 p_x \mathbbm{p} = 0 , \label{pde:1}\\
 &D_t \mathbbm{v}_{0} + \partial _{x} \mathbbm{v}_{1} = 0 , \label{pde:2}\\
 &D_t \mathbbm{v}_{1} + \partial _{x} \mathbbm{v}_{0} = -2 m \mathbbm{p} , \label{pde:3}\\
 &D_t \mathbbm{p} + 2 p_x \mathbbm{s} = 2 m \mathbbm{v}_{1} , \label{pde:4}
\end{align}
with pseudodifferential operator
\begin{equation}\label{pseudoDiff}
 D_t = \partial_{t} + e \int_{-1/2}^{1/2} d \xi \,\,\, E_{x} \left( x + i \xi \partial_{p_{x}} \, , t \right) \partial_{p_{x}} .
\end{equation}
For the brevity of the symbol notation, we define Wigner components $\mathbbm{w}_{0} = \mathbbm{s}$ , $\mathbbm{w}_{1} = \mathbbm{v}_{0}$ , $\mathbbm{w}_{2} = \mathbbm{v}_{1}$ and $\mathbbm{w}_{3} = \mathbbm{p}$.
In order to perform the calculation, it is also necessary to choose vacuum initial condition values, we employ them \cite{Kohlfurst:2015zxi}
\begin{equation}\label{vacuum-initial}
{\mathbbm{w}}_{0 \, \rm{vac}} = - \frac{2m}{\Omega} \, ,
\quad  {\mathbbm{w}}_{2 \, \rm{vac}} = - \frac{2{ p_x} }{\Omega} \,  ,
\end{equation}
where $\Omega$ is the energy of single particle and defined as $\Omega=\sqrt{p_{x}^{2}+m^2}$. By subtracting these initial vacuum terms, we can obtain modified Wigner components
\begin{equation}
\mathbbm{w}_{k}^{v} \left( x , p_{x} , t \right) = \mathbbm{w}_{k} \left( x , p_{x} , t \right) - \mathbbm{w}_{{k} \, \rm{vac}}\left(p_{x} \right),
\end{equation}
where $\mathbbm{w}_{k}$ is the Wigner component in Eqs. \eqref{pde:1}-\eqref{pde:4}, and $\mathbbm{w}_{{k} \, \rm{vac}}$ is the corresponding vacuum initial condition Eq.\eqref{vacuum-initial}.
Total energy of the Dirac particle is
$\varepsilon \left( x , p_{x} , t \right) = {m  \mathbbm{s}^{v} \left( x , p_{x} , t \right) + p_{x}  \mathbbm{v}_{1}^{v} \left( x , p_{x} , t \right)}$,
and the particle number density in phase space can be written as \cite{Hebenstreit:2011wk}
\begin{equation}\label{particle number density}
n \left( x , p_{x} , t \right) = \frac{m  \mathbbm{s}^{v} \left( x , p_{x} , t \right) + p_{x}  \mathbbm{v}_{1}^{v} \left( x , p_{x} , t \right)}{\Omega \left( p_{x} \right)},
\end{equation}
where ${\Omega \left( p_{x} \right)}$ is energy of the individual particle. Therefore, the momentum distribution function can be obtained by integration of Eq. \eqref{particle number density} with respect to position
\begin{equation}\label{momentum distribution}
n\left( p_x,t \right) = \int dx \, n \left( x , p_{x} , t \right).
\end{equation}
Also the total number of created particles could be written
 \begin{equation}\label{Num}
N\left(t \right) = \int dx\,dp_x n \left( x , p_{x} , t \right).
\end{equation}
Moreover, in order to obtain the nontrivial spatial dependence of results on $\lambda$, we calculate the reduced quantities $\bar{n}\left( p_{x}, t \right)\equiv n\left( p_{x}, t \right)/\lambda$ and $\bar{N}\left(t\rightarrow\infty\right)/{\lambda} \equiv \bar{N}$.

\section{Numerical results }\label{results}
In this paper, we study the symmetrical frequency chirp effects for both high frequency and low frequency inhomogeneous fields. We consider two carrier phase parameters $\varphi=0$ and $\varphi=\pi/2$ representing cosine- and sine-type oscillating forms respectively. For the convenience of comparison, we choose similar parameter values as in Ref. \cite{Ababekri:2020} where the usual asymmetrical chirp is studied.

\subsection{High frequency field}\label{result1}

For high frequency field, we set $\omega=0.7m$, $\tau=45m^{-1}$ for which the electron-positron pair is predominantly produced by multiphoton absorbtion.
For $b=0$, the external field has a constant frequency $\omega_{\text{eff}}(t)=\omega$ and we would recover the result in Fig.~1 of Ref.~\cite{Ababekri:2020} where the decreasing spatial scale affects the discrete $n$-photon absorption spectrum ~\cite{Kohlfurst:2017hbd,Ababekri:2020}.

\subsubsection{Momentum distribution when $\varphi=0$}

\begin{figure}[H]
\begin{center}
\includegraphics[width=\textwidth]{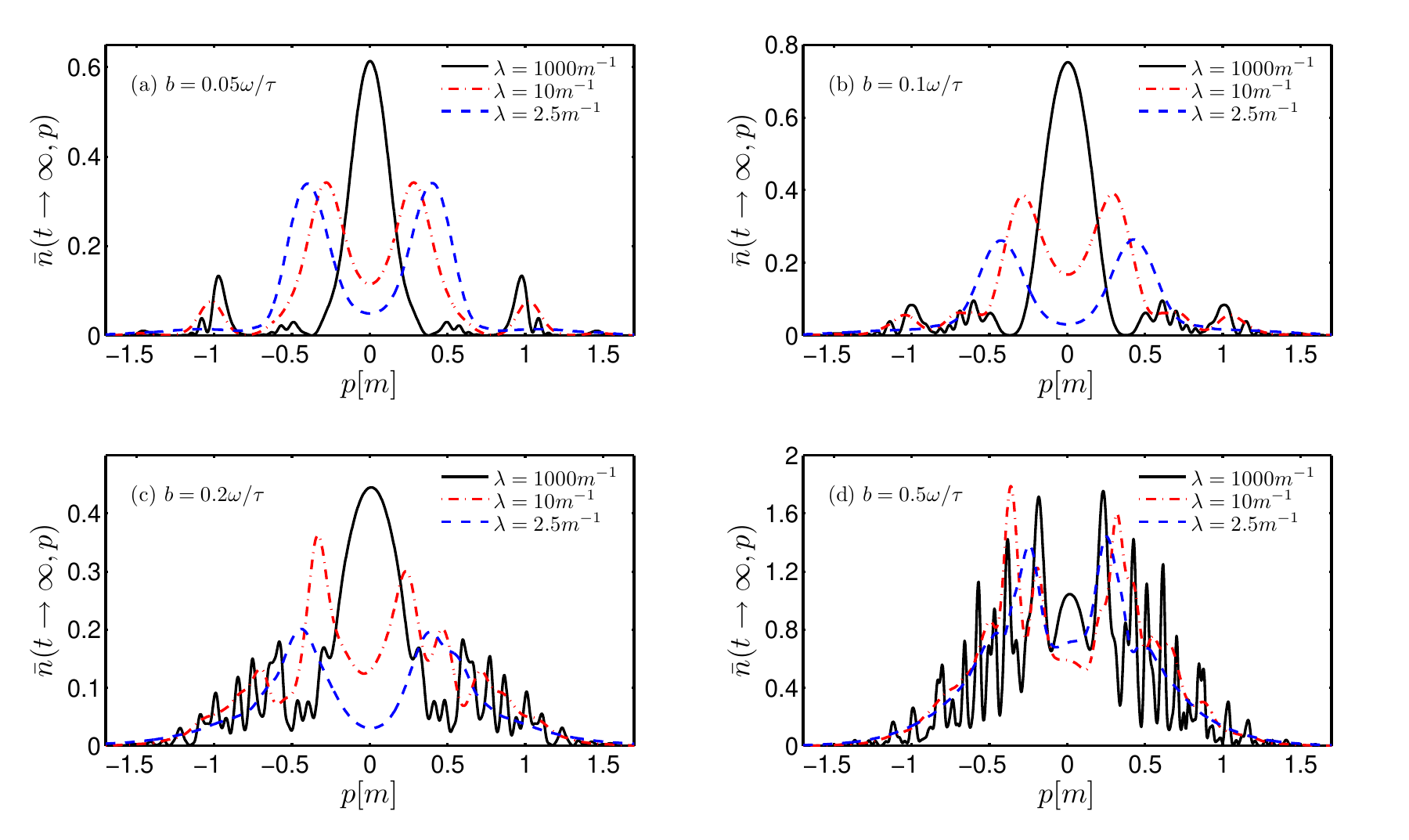}
\end{center}
  \caption{(color online). Reduced momentum spectrum for various chirp values in high frequency field with different spatial scales. The chirp values are $b=0.05\omega/\tau \approx 0.00078 m^2$, $b=0.1\omega/\tau \approx 0.0016 m^2$, $b=0.2\omega/\tau \approx 0.0031 m^2$ and $b=0.5\omega/\tau \approx 0.0078 m^2$. Other field parameters are $\epsilon=0.5$, $\omega=0.7 m$, $\tau=45m^{-1}$ and $\varphi=0$.}
  \label{omg07pi0}
\end{figure}

The symmetrical chirp effects on momentum spectrum at different spatial scales are shown in Fig.~\ref{omg07pi0}.
At the large spatial scale, $\lambda=1000m^{-1}$, the momentum spectrum is very sensitive to changing of chirp parameter. Weak oscillation can be observed even for small chirp, see Fig.~\ref{omg07pi0}(a) and (b). Strong oscillation occurs in the momentum spectrum as chirp increases, see Fig.~\ref{omg07pi0}(c) and (d). For nonzero chirp, the effective frequency $\omega_{\text{eff}}(t)=\omega+b|t|$ increases with chirp and provides more energy to create more particles. Consequently, these particles interact with each other can lead the more obvious interference effect on the momentum spectrum, this result is similar to that of homogeneous field.

At the small spatial scales, $\lambda=2.5m^{-1}$ and $\lambda=10m^{-1}$, for small chirp, the oscillation is not obvious, but we observed peak splitting on momentum spectrum, see Fig.~\ref{omg07pi0}(a) and (b). The peak splitting can be explained by ponderomotive effects which have been reported in Ref. \cite{Kohlfurst:2017hbd}. The magnitude of ponderomotive force is inversely proportional to the size of spatial scale \cite{Kohlfurst:2015zxi}. Therefore the momentum peaks at spatial scale $\lambda=2.5m^{-1}$ are more far away from the center compared with the case of $\lambda=10m^{-1}$ in Fig.~\ref{omg07pi0}(a) and (b). However, for large chirp, the peaks that split by the ponderomotive force are replaced by strong oscillation as shown in Fig.~\ref{omg07pi0}(c) and (d). Because the effective frequency increases obviously with the large chirp and provides greater contribution to frequency of external field. So that corresponding effect of ponderomotive force decreases and strong oscillation occurs on the momentum spectrum. This oscillation can be understood as the interference effects between created particles.

\begin{figure}[H]
\begin{center}
\includegraphics[scale=0.55]{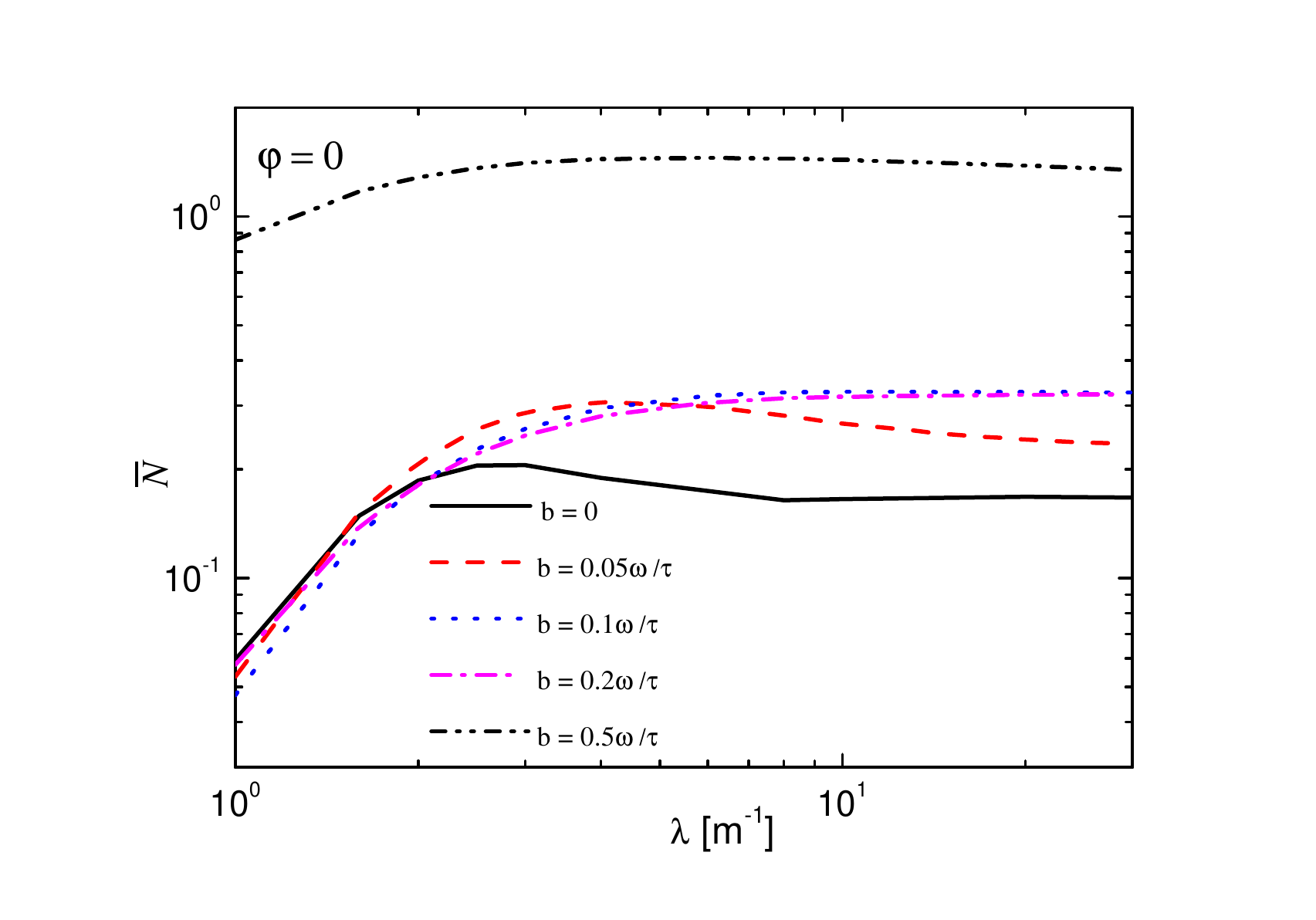}
\end{center}
\caption{(color online). Reduced particle number dependence on spatial scales for various chirp values in high frequency field. Other field parameters are the same as in Fig.~\ref{omg07pi0}.}
\label{omg07particle nummber}
\end{figure}

In Fig.~\ref{omg07particle nummber}, we also plot the reduced particle number dependence on spatial scales for various chirp parameters. We can see that reduced particle number is enhanced with the increasing chirp. This enhancement is most pronounced for the largest chirp, see in Fig.~\ref{omg07particle nummber}. When chirp is relatively small, i.e., $b \leq0.2\omega/\tau \approx 0.0031 m^2$, reduced particle number is enhanced rapidly as spatial scale increases. Because the electric field energy increases with widening of spatial scale, particles created in the field region also increase correspondingly. However, when chirp is very large, i.e., $b=0.5\omega/\tau \approx 0.0078 m^2$, particle number is less affected by spatial scales. It exhibits a relative flat line in a whole region of spatial scale of field. At large spatial scales, particle number tends to be a constant for each chirp parameters as spatial scale increases. At that time, our field can be regarded as a spatial homogeneous field. We noted that, compared with the case of $b=0$ (solid line), reduced particle number is enhanced by one order of magnitude for largest chirp (dash-dot-dotted line).

For a convenient comparison between our results to that of the field with asymmetrical frequency chirping case in Ref. \cite{Ababekri:2020}, it is intuitive to see their difference for various chirp parameters at fixed spatial scale, meanwhile, we can also see their difference for various spatial scales with fixed chirp parameter. As two examples by fixed $\lambda=10m^{-1}$ and fixed $b=0.5\omega/\tau=0.0078m^2$, the comparison data of reduced particle numbers for our case and asymmetrical frequency chirping case is presented in Table~\ref{Table 1}.

\begin{table}[H]
\caption{The reduced particle numbers of symmetrical chirp in present study and asymmetrical chirp in Ref. \cite{Ababekri:2020} and the ratio of them, $\bar{N}_{sym}(\varphi=0)$/$\bar{N}_{asym}(\varphi=0)$, for different $b$ when $\lambda=10m^{-1}$ is fixed (the upper part of table), and for different $\lambda$ when $b=0.5\omega/\tau=0.0078m^2$ is fixed (the lower part of table). Field parameters are $\epsilon=0.5$, $\omega=0.7m$ and $\tau=45m^{-1}$.}
\centering
\begin{ruledtabular}
\begin{tabular}{cccc}
$b$ &$\bar{N}_{sym}(\varphi=0)$    &$\bar{N}_{asym}(\varphi=0)$ &$\bar{N}_{sym}(\varphi=0)$/$\bar{N}_{asym}(\varphi=0)$\\
\hline
$0.05\omega/\tau \approx0.00078m^{2}$   &$0.268$       & $0.1663$            & $1.612$\\
$0.1\omega/\tau \approx0.0016m^{2}$     &$0.3274$     &$ 0.1561$             & $2.097$\\
$0.2\omega/\tau \approx0.0031m^{2}$    &$0.3178$     &$ 0.1791$             & $1.774$\\
$0.5\omega/\tau \approx0.0078m^{2}$    &$1.434$     &$ 0.7474$              & $1.919$\\
\hline
$\lambda$ &$\bar{N}_{sym}(\varphi=0)$    &$\bar{N}_{asym}(\varphi=0)$ &$\bar{N}_{sym}(\varphi=0)/\bar{N}_{asym}(\varphi=0)$\\
\hline
$2.5m^{-1}$      &$1.3639$                &$0.7059$                      & $1.93$\\
$10m^{-1}$     &$1.434$                 &$0.7474$                       & $1.919$\\
$1000m^{-1}$    &$1.2601$                 &$0.7268$                       & $1.73$\\
\end{tabular}
\end{ruledtabular}
\label{Table 1}
\end{table}

Compared to the case of asymmetrical frequency chirp, the reduced particle number is enhanced at about two times in our symmetrical chirp case. The enhancement is most pronounced for $b=0.1\omega/\tau \approx0.0016m^{2}$ and $\lambda=2.5m^{-1}$.

\subsubsection{Momentum distribution when $\varphi=\pi/2$}\label{resultB}

\begin{figure}[H]
\begin{center}
\includegraphics[width=\textwidth]{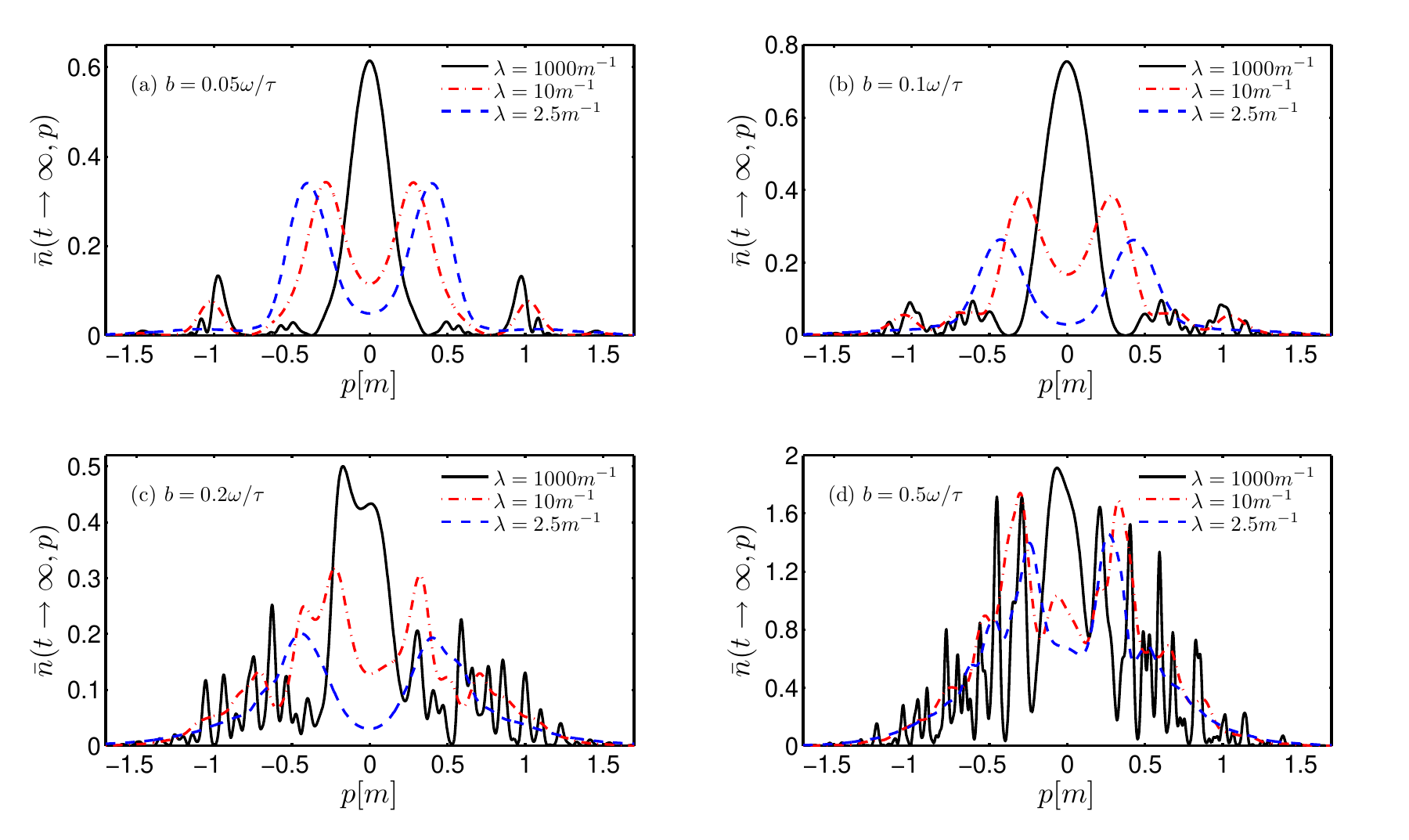}
\end{center}
  \caption{(color online). Reduced momentum spectrum for various chirp parameters in the high frequency field with different spatial scales. The field parameters are the same as in Fig. \ref{omg07pi0} except $\varphi=\pi/2$.}
  \label{omg07pi05}
\end{figure}

When $\varphi=\pi/2$, we show the symmetrical chirp effects on momentum spectrum at different spatial scales in Fig.~\ref{omg07pi05}.
For small chirp, see Fig.~\ref{omg07pi05}(a) and (b), we obtain almost the same results in both phase cases, see also Fig.~\ref{omg07pi0}(a) and (b). At large spatial scale $\lambda=1000m^{-1}$, however, some delicate differences can be observed on momentum spectrum when $b$ is large, see Fig.~\ref{omg07pi05}(c) and (d) in comparison with the results of $\varphi=0$ in Fig.~\ref{omg07pi0}(c) and (d). For example, when $b=0.2\omega/\tau \approx 0.0031 m^2$, the main peak value corresponding to $p=0$ is split by strong oscillation, see Fig.~\ref{omg07pi05}(c). Moreover it reaches larger peak value $\bar{n}(t\rightarrow\infty, p)=1.9m$ for $b=0.5\omega/\tau \approx 0.0078 m^2$, see Fig.~\ref{omg07pi05}(d). On the other hand, however, at small spatial scales $\lambda=2.5m^{-1}$ and $\lambda=10m^{-1}$, while we can still observe oscillation on the momentum spectrum, there is no obvious increasing of peak values compared with the former case of $\varphi=0$.

\begin{figure}[H]
\begin{center}
\includegraphics[scale=0.5]{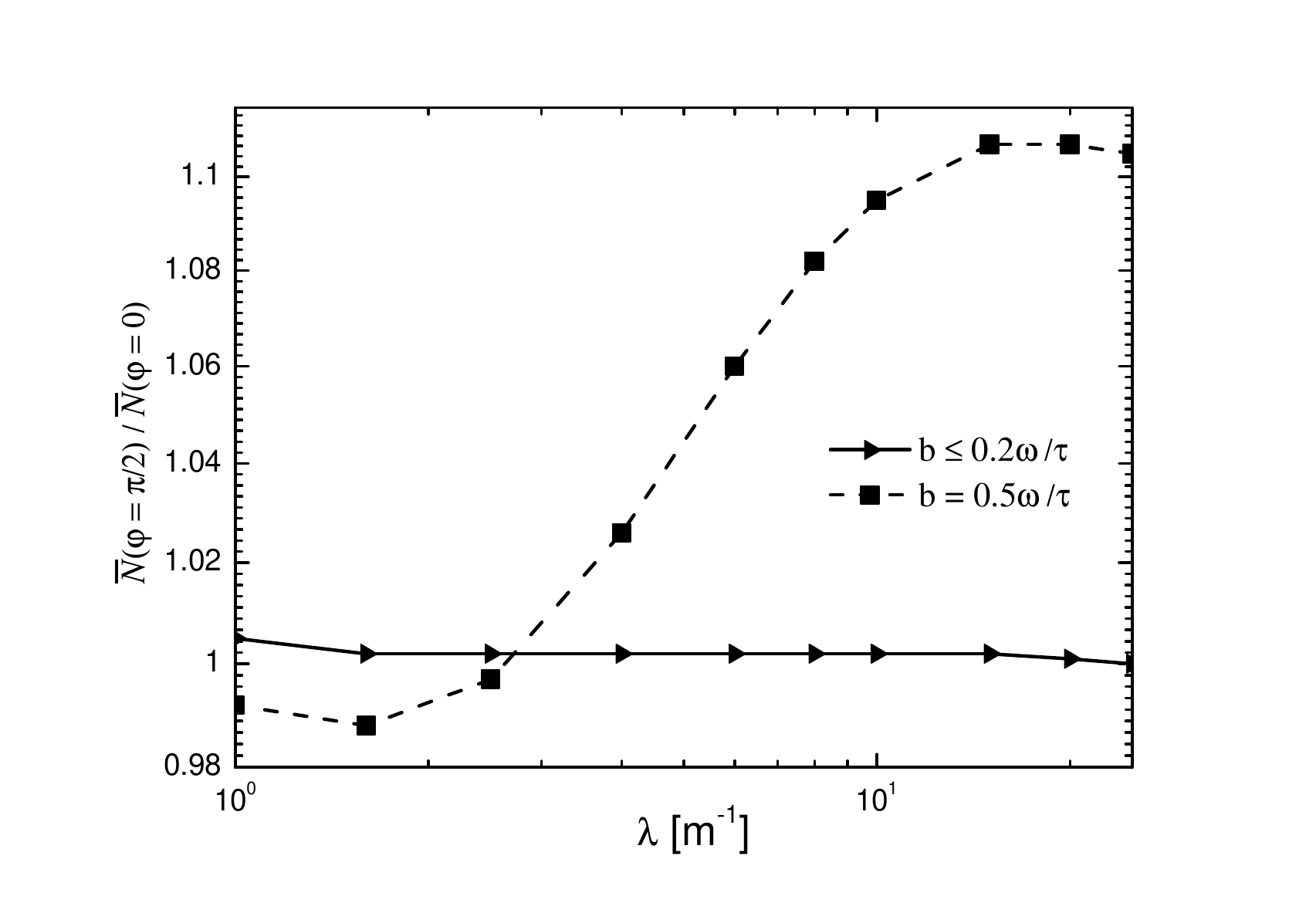}
\end{center}
\caption{The ratio of reduced particle numbers $\bar{N}(\varphi=\pi/2)$/$\bar{N}(\varphi=0)$ for different cases of chirp $b \leq 0.2\omega/\tau \approx 0.0031m^2$ (solid line) and $b=0.5\omega/\tau \approx 0.0078m^2$ (dashed line).}
\label{omg07ratio}
\end{figure}

The ratio of reduced particle numbers $\bar{N}(\varphi=\pi/2)$/$\bar{N}(\varphi=0)$ for different cases of chirp $b \leq 0.2\omega/\tau \approx 0.0031m^2$ and $b=0.5\omega/\tau \approx 0.0078m^2$ are shown in Fig.~\ref{omg07ratio}. It is found that when $b \leq 0.2\omega/\tau \approx 0.0031m^2$, the ratio for different chirp is almost equal to $1$, see the solid line of figure. We have checked it to be valid at different spatial scales for chirp when $b=0$, $b=0.05\omega/\tau \approx 0.00078m^2$, $b=0.1\omega/\tau \approx 0.0016m^2$ as well as $b=0.2\omega/\tau \approx 0.0031m^2$. On the other hand, however, when $b=0.5\omega/\tau \approx 0.0078m^2$, the particle number $\bar{N}(\varphi=\pi/2)$ is enhanced a little as spatial scale increases.

The asymmetrical chirp effect when $\varphi=\pi/2$ has not been considered in Ref. \cite{Ababekri:2020}. For comparison, we also perform some numerical calculations in the asymmetrical chirping field. The comparison data of reduced particle numbers when $\varphi=\pi/2$ is given in Table~\ref{Table 2}.
We find that compared to the case of asymmetrical frequency chirp, the reduced particle number is enhanced at about two times in our symmetrical chirping case. The largest enhancement still occurs for $b=0.1\omega/\tau \approx0.0016m^{2}$ and $\lambda=2.5m^{-1}$.

\begin{table}[H]
\caption{The reduced particle numbers of symmetrical chirp in present study and asymmetrical chirp in Ref. \cite{Ababekri:2020} and the ratio of them, $\bar{N}_{sym}(\varphi=\pi/2)$/$\bar{N}_{asym}(\varphi=\pi/2)$, for different $b$ when $\lambda=10m^{-1}$ is fixed (the upper part of table), and for different $\lambda$ when $b=0.5\omega/\tau=0.0078m^2$ is fixed (the lower part of table). Other field parameters are the same in Table~\ref{Table 1}.}
\centering
\begin{ruledtabular}
\begin{tabular}{cccc}
$b$ &$\bar{N}_{sym}(\varphi=\pi/2)$    &$\bar{N}_{asym}(\varphi=\pi/2)$ &$\bar{N}_{sym}(\varphi=\pi/2)$/$\bar{N}_{asym}(\varphi=\pi/2)$\\
\hline
$0.05\omega/\tau \approx0.00078m^{2}$   &$0.2682$       & $0.1663$            & $1.61$\\
$0.1\omega/\tau \approx0.0016m^{2}$     &$0.3279$     &$ 0.1561$             & $2.1$\\
$0.2\omega/\tau \approx0.0031m^{2}$    &$0.3185$     &$ 0.1773$             & $1.8$\\
$0.5\omega/\tau \approx0.0078m^{2}$    &$1.5702$     &$ 0.8304$              & $1.89$\\
\hline
$\lambda$ &$\bar{N}_{sym}(\varphi=\pi/2)$    &$\bar{N}_{asym}(\varphi=\pi/2)$ &$\bar{N}_{sym}(\varphi=\pi/2)/\bar{N}_{asym}(\varphi=\pi/2)$\\
\hline
$2.5m^{-1}$      &$1.3596$                &$0.7085$                           & $1.92$\\
$10m^{-1}$     &$1.5702$                 &$0.8304$                       & $1.89$\\
$1000m^{-1}$    &$1.3825$                 &$0.8124$                       & $1.7$\\
\end{tabular}
\end{ruledtabular}
\label{Table 2}
\end{table}

\subsection{low frequency field}\label{result2}

In this section, we choose $\omega=0.1m$ and $\tau=25m^{-1}$ to study symmetrical chirp effects on the momentum spectrum and reduced particle number in low frequency fields with $\varphi=0$ and $ \varphi=\pi/2$.

\subsubsection{Momentum distribution when $\varphi=0$}\label{resultC}

\begin{figure}[h]
\begin{center}
\includegraphics[width=\textwidth]{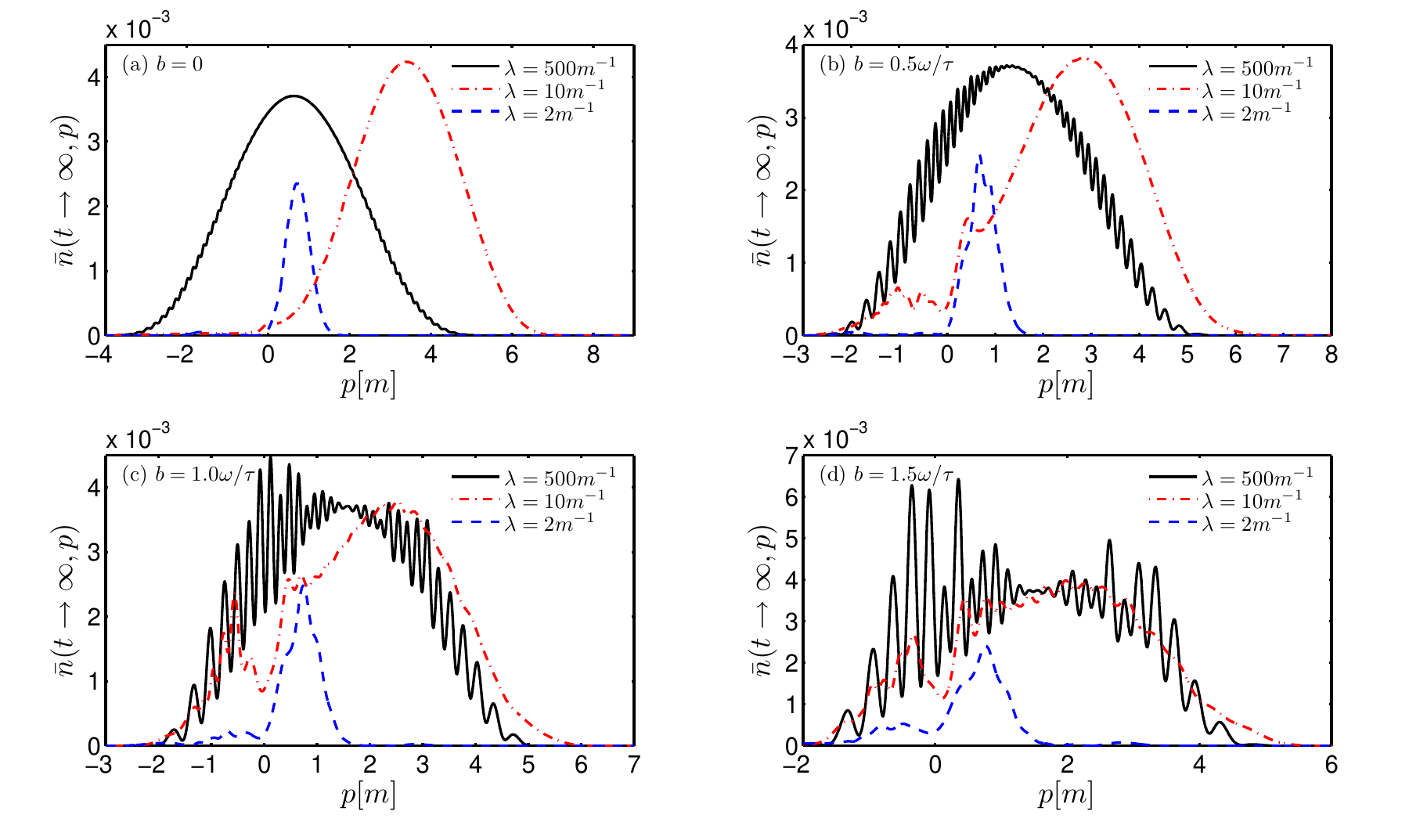}
\end{center}
  \caption{(color online). Reduced momentum spectrum for various chirp parameters in the low frequency field with different spatial scales. The chirp values are $b=0$, $b=0.5\omega/\tau=0.002m^2$, $b=1.0\omega/\tau=0.004m^2$ and $b=1.5\omega/\tau=0.006m^2$. Other field parameters are $\epsilon=0.5$, $\omega=0.1m$, $\tau=25m^{-1}$ and $\varphi=0$.}
  \label{omg01pi0}
\end{figure}
When $b=0$, see Fig.~\ref{omg01pi0}(a), momentum spectrum is same as in Fig. 6(a) of Ref. \cite{Ababekri:2020}. At large spatial scale $\lambda=500m^{-1}$, when $b=0$ , a very weak oscillation can be observed. For small chirp, we can see stronger oscillation on the two sides of momentum spectrum in Fig.~\ref{omg01pi0}(b). For the large chirp, very strong oscillation appears, meanwhile, momentum spectrum shrinks and shifts towards positive direction, see Fig.~\ref{omg01pi0}(c) and (d). These oscillations can be understood as interference effect of particles created by opposite signed large peaks of the temporal field \cite{Hebenstreit:2009km}.

When the spatial scale is narrowed to $\lambda=10m^{-1}$, when $b=0$, it has no oscillation on the momentum spectrum. For small chirp, we can observed a weak oscillation on the left side, meanwhile, momentum spectrum is broadened to negative field region, see dash-dotted line in Fig.~\ref{omg01pi0}(b). This is because particles created with certain momentum leave the field region and miss the deceleration by the negative field peak \cite{Ababekri:2020}. For large chirp, the strong oscillation occurs on momentum spectrum, it is caused by the interference effect of created particles from opposite field peaks, see Fig.~\ref{omg01pi0}(c) and (d).

When $\lambda=2m^{-1}$, for small chirp, we have not observed obvious oscillation on the momentum spectrum. For large chirp, a weak oscillation appeared, see dashed line in Fig.~\ref{omg01pi0}(c) and (d). Because, the spatial scale is very small, the work done by the electric field is reduced correspondingly. Consequently, particles created by the electric field are also decreased, so that no pronounced interference can be observed on momentum spectrum.

In the low frequency field, we also make a comparison between the reduced particle number with our case and that with asymmetrical frequency chirp. We show the comparison data of reduced particle numbers in Table~\ref{Table 3}. Compared to the case of asymmetrical frequency chirp, the reduced particle number is almost unchanged in our symmetrical chirping case.

\begin{table}[H]
\caption{The reduced particle numbers of symmetrical chirp in present study and asymmetrical chirp in Ref. \cite{Ababekri:2020} and the ratio of them, $\bar{N}_{sym}(\varphi=0)$/$\bar{N}_{asym}(\varphi=0)$, for different $b$ when $\lambda=2m^{-1}$ is fixed (the upper part of table), and for different $\lambda$ when $b=0.5\omega/\tau=0.002m^2$ is fixed (the lower part of table). Field parameters are $\epsilon=0.5$, $\omega=0.1m$, and $\tau=25m^{-1}$.}
\centering
\begin{ruledtabular}
\begin{tabular}{cccc}
$b$ &$\bar{N}_{sym}(\varphi=0)$ &$\bar{N}_{asym}(\varphi=0)$ &$\bar{N}_{sym}(\varphi=0)/\bar{N}_{asym}(\varphi=0)$\\
\hline
$0$                            &$0.0018$      &$0.0018$            & $1$\\
$0.5\omega/\tau=0.002m^{2}$     &$0.0018$      &$0.0018$             & $1$\\
$1.0\omega/\tau=0.004m^{2}$    &$0.0020$      &$0.0019$             & $1.052$\\
$1.5\omega/\tau=0.006m^{2}$    &$1.0025$       &$0.0022$             & $1.136$\\
\hline
$\lambda$ &$\bar{N}_{sym}(\varphi=0)$    &$\bar{N}_{asym}(\varphi=0)$ &$\bar{N}_{sym}(\varphi=0)/\bar{N}_{asym}(\varphi=0)$\\
\hline
$2m^{-1}$      &$0.0018$                &$0.0018$                           & $1$\\
$10m^{-1}$     &$0.0135$                 &$0.0140$                       & $0.96$\\
$500m^{-1}$    &$0.0144$                 &$0.0149$                       & $0.97$\\
\end{tabular}
\end{ruledtabular}
\label{Table 3}
\end{table}

\subsubsection{Momentum distribution when $\varphi=\pi/2$}\label{resultD}

\begin{figure}[h]
\begin{center}
\includegraphics[width=\textwidth]{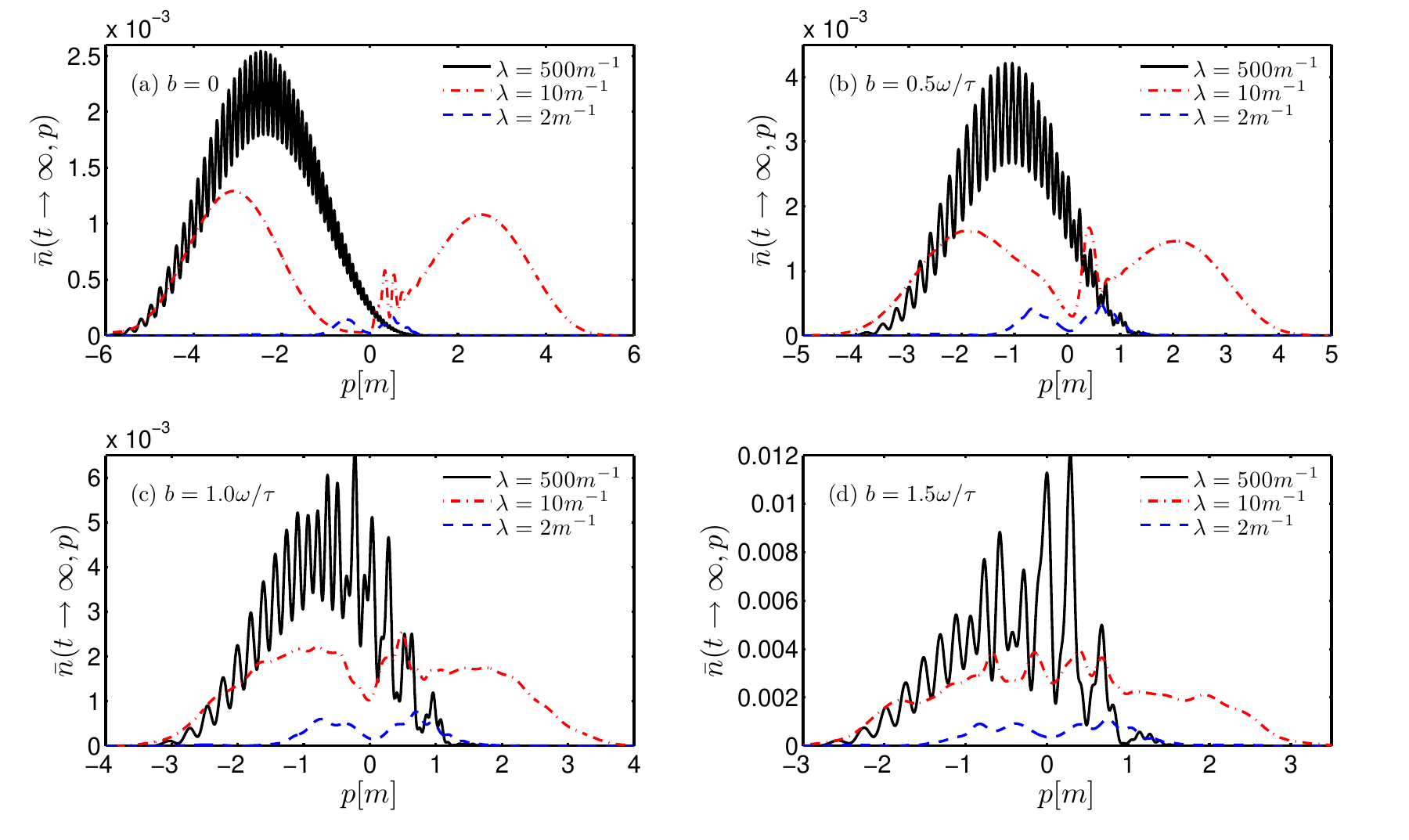}
\end{center}
  \caption{(color online). Reduced momentum spectrum for various chirp values in the low frequency field with different spatial scales. The field parameters are the same as in Fig. \ref{omg01pi0} except $\varphi=\pi/2$.}
  \label{omg01pi05}
\end{figure}

The symmetrical chirp effects on momentum spectrum at different spatial scales with $\varphi=\pi/2$ are shown in Fig.~\ref{omg01pi05} where the remarkable difference in the momentum spectrum compared with that of $\varphi=0$ shown in Fig.~\ref{omg01pi0} could be noticed. When $b=0$, see Fig.~\ref{omg01pi05}(a), momentum spectrum is same as in Fig. 8(a) of Ref. \cite{Ababekri:2020}.

At large spatial scale $\lambda=500m^{-1}$, when $b=0$, we can see the bell-shaped momentum spectrum in the negative region with oscillation which can be understood as the interference effect between two separated pair production events that happen at different times \cite{Hebenstreit:2009km}, see Fig.~\ref{omg01pi05}(a). For small chirp, momentum spectrum shrinks and shifts towards positive direction, see solid line in Fig.~\ref{omg01pi05}(b). For large chirp, the bell-shaped momentum spectrum is replaced by strong oscillation, see Fig. \ref{omg01pi05}(c) and (d). In the next section, we will discuss some of these oscillation effects in the semiclassical picture.

When the spatial scale is decreased to $\lambda=10m^{-1}$, chirp effects on momentum spectrum become stronger than that of $\varphi=0$. For $b=0$, pair is created by two opposite signed field peaks and leave field region in two directions \cite{Ababekri:2020}, so that form the two peaks on momentum spectrum, see Fig.~\ref{omg01pi05}(a) which is same as Fig. 8(a) of Ref. \cite{Ababekri:2020}. For small chirp, i.e., $b=0.5\omega/\tau=0.002m^2$, the two momentum peaks are still remained, it is different from the result for asymmetrical chirped field where one peak vanishes when the same $b=0.5\omega/\tau=0.002m^2$ is applied, see Fig. 8(b) in Ref. \cite{Ababekri:2020}. However, for large chirp, these peaks are replaced by interference pattern of created particles, see Fig.~\ref{omg01pi05}(c) and (d).

At the small spatial scale $\lambda=2m^{-1}$, two momentum peaks can also be observed, but the peak values are very small compared with the case of $\lambda=10m^{-1}$. We think it is because that, in the small spatial scale $\lambda=2m^{-1}$, the work done by the electric field would be so greatly reduced and very few of the particles are created through tunneling.

\begin{figure}[H]
\begin{center}
\includegraphics[scale=0.55]{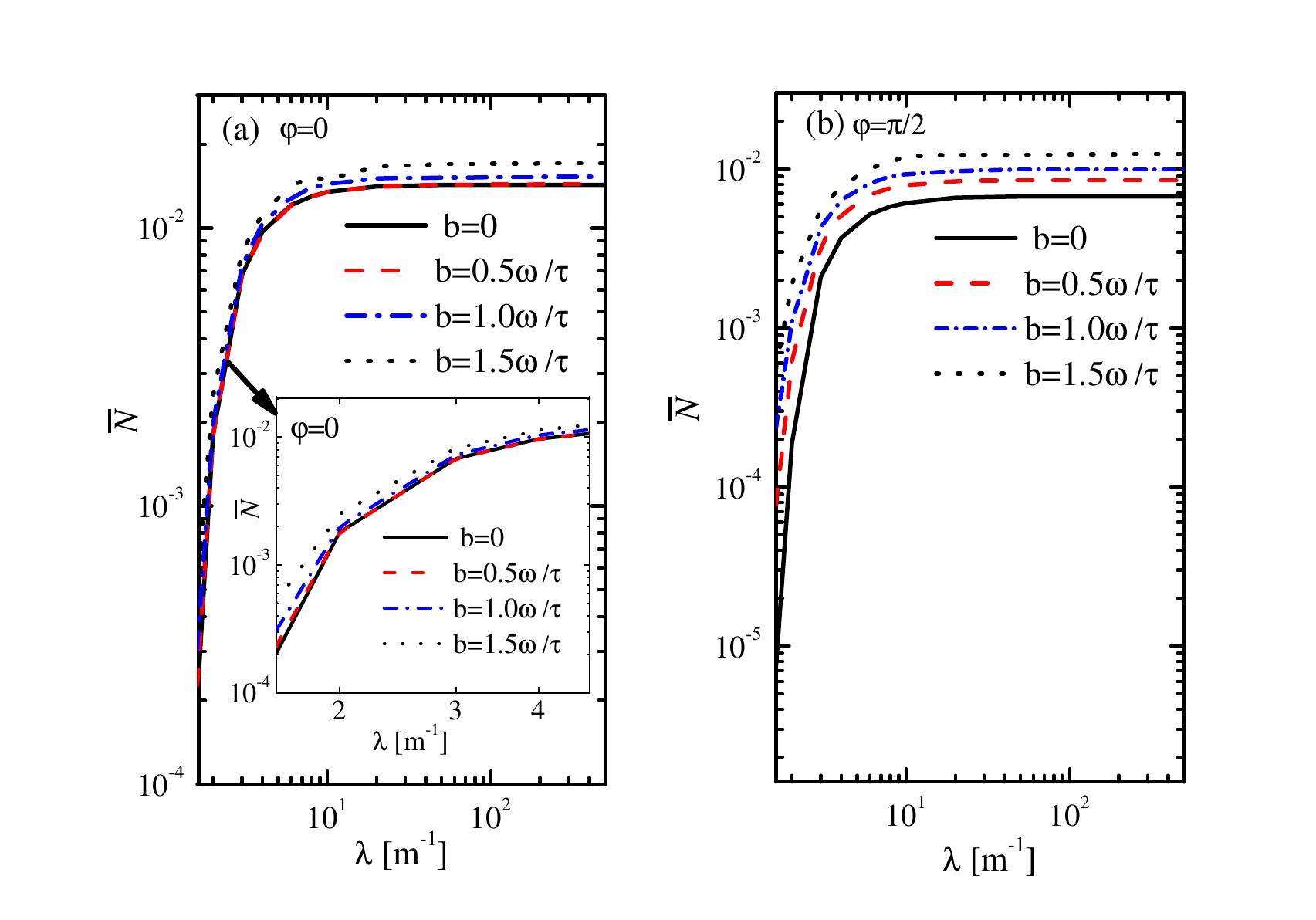}
\end{center}
\caption{(color online). Reduced particle number dependence on spatial scales for various chirp parameters in the low frequency field with different carrier phase parameter. The carrier phase parameters are $\varphi=0$ for (a) and $\varphi=\pi/2$ for (b), respectively. Other field parameters are the same as in Fig. \ref{omg01pi0}.}
\label{omg01particle nummber}
\end{figure}

Now let us to see the reduced particle number dependence on spatial scales for various chirp parameters, shown in the Fig.~\ref{omg01particle nummber}.
When $\varphi=0$, the reduced particle number is not sensitive to small chirp. For small chirp $b=0.5\omega/\tau$, for instance, the created particle number is almost similar to the case of $b=0$ without chirping, see the solid and dashed lines in Fig.~\ref{omg01particle nummber}(a). However the reduced particle number enhancement occurs for large chirp. Moreover, when $\varphi=\pi/2$, compared with the case of $\varphi=0$, particle number enhancement is more sensitive to increasing chirp. We can observe an obvious enhancement on the reduced particle number even for the small chirp, see Fig.~\ref{omg01particle nummber}(b). At small spatial scales, for both of two cases, the reduced particle number is enhanced rapidly as spatial scale enlarges. Compared to the case of $b=0$ without chirping, the reduced particle number is increased by two orders of magnitude for largest chirp when $\varphi=\pi/2$, see solid and dotted line in Fig.~\ref{omg01particle nummber}(b). At large spatial scale, reduced particle number tends to be a constant no matter how spatial scale enlarges.
Furthermore, for the same chirp, the reduced particle number in the field with carrier phase $\varphi=\pi/2$ is always smaller than that of the field with $\varphi=0$.

When $\varphi=\pi/2$, we also compare our results to that of asymmetrical frequency chirp in Ref. \cite{Ababekri:2020} and the comparison data of reduced particle numbers for two cases is shown in Table~\ref{Table 4}.

\begin{table}[H]
\caption{The reduced particle numbers of symmetrical chirp in present study and asymmetrical chirp in Ref. \cite{Ababekri:2020} and the ratio of them, $\bar{N}_{sym}(\varphi=\pi/2)$/$\bar{N}_{asym}(\varphi=\pi/2)$, for different $b$ when $\lambda=2m^{-1}$ is fixed (the upper part of table), and for different $\lambda$ when $b=0.5\omega/\tau=0.002m^2$ is fixed (the lower part of table). Other field parameters are the same in Table~\ref{Table 3}.}
\centering
\begin{ruledtabular}
\begin{tabular}{cccc}
$b$ &$\bar{N}_{sym}(\varphi=\pi/2)$ &$\bar{N}_{asym}(\varphi=\pi/2)$ &$\bar{N}_{sym}(\varphi=\pi/2/\bar{N}_{asym}(\varphi=\pi/2)$\\
\hline
$0$                            &$1.88\times10^{-4}$      &$1.88\times10^{-4}$            & $1$\\
$0.5\omega/\tau=0.002m^{2}$     &$6.26\times10^{-4}$      &$3.15\times10^{-4}$             & $1.987$\\
$1.0\omega/\tau=0.004m^{2}$    &$1.1\times10^{-3}$      &$5.61\times10^{-4}$             & $1.96$\\
$1.5\omega/\tau=0.006m^{2}$    &$1.9\times10^{-3}$       &$9.26\times10^{-4}$             & $2.05$\\
\hline
$\lambda$ &$\bar{N}_{sym}(\varphi=\pi/2)$    &$\bar{N}_{asym}(\varphi=\pi/2)$ &$\bar{N}_{sym}(\varphi=\pi/2)/\bar{N}_{asym}(\varphi=\pi/2)$\\
\hline
$2m^{-1}$      &$6.26\times10^{-4}$      &$3.15\times10^{-4}$            & $1.987$\\
$10m^{-1}$     &$0.0079$                 &$0.0049$                       & $1.612$\\
$500m^{-1}$    &$0.0085$                 &$0.0053$                       & $1.603$\\
\end{tabular}
\end{ruledtabular}
\label{Table 4}
\end{table}

This table clearly demonstrates that the reduced particle number in our symmetrical frequency chirping field is increased about two times of asymmetrical frequency chirp case. The largest enhancement occurs for $b=1.5\omega/\tau=0.006m^2$ and $\lambda=2m^{-1}$.

\section{Discussions}\label{Discussions}

\begin{figure}[H]
  \centering
  \subfigure{\includegraphics[width=4.5cm]{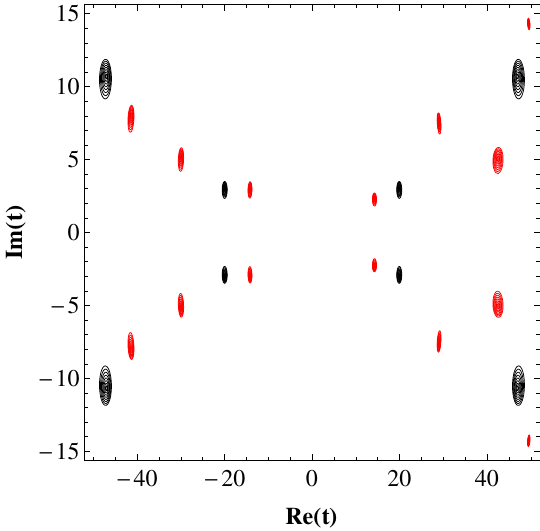}}
  \hspace{7mm}
  \subfigure{\includegraphics[width=4.5cm]{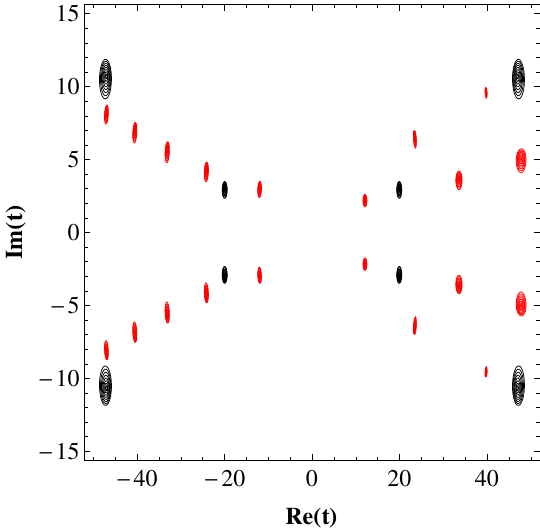}}
  \hspace{7mm}
  \subfigure{\includegraphics[width=4.5cm]{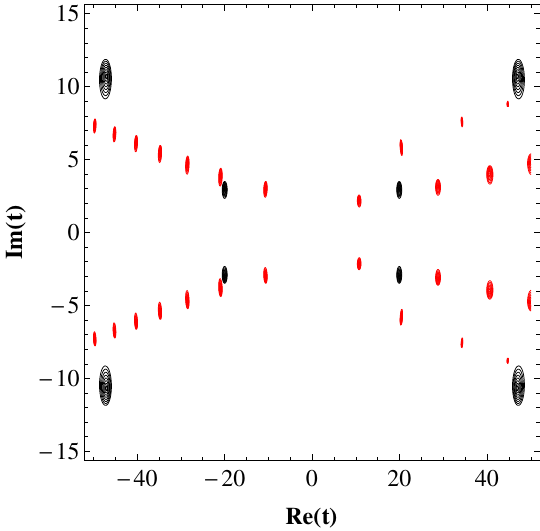}}
  \caption{(color online). Contour plots of $|\omega_{\bf p}(t)|^2$ in the complex $t$ plane, showing the turning point distribution where $\omega_{\bf p}(t)=0$ for $p_{\perp}=0$, $p_{x}=0$. The black dots are turning point for the case of $b=0$ without chirping, the red dots from left to right are the complex conjugate turning points for various chirp parameter: $b=0.5\omega/\tau=0.002m^2$(left panel), $b=1.0\omega/\tau=0.004m^2$(center panel) and $b=1.5\omega/\tau=0.006m^2$(right panel), respectively. Other field parameters are $\epsilon=0.5$, $\omega=0.1 m$, $\tau=25m^{-1}$ and $\varphi=\pi/2$. }
\label{turning point}
\end{figure}

The symmetrical chirp effects on the momentum spectrum and reduced particle number can also be discussed in terms of the turning point structure of potential in semiclassical regime by employing complex Wentzel-Kramers-Brillouin (WKB) scattering approach. The scattering potential is determined by solving the equation $\omega_{\bf p}(t)= \sqrt{m^{2}+p_{\perp}^{2}+(p_{x}-eA(t))^{2}}=0$ for the turning points which appear as complex conjugate pairs when the potential is real \cite{Dumlu:2010vv}. When a single complex turning point dominates, the WKB results for the particle creation rate is given as \cite{heading,landau}
\begin{equation}
N_{\rm}\approx e^{-2K} , \qquad K=\left| \int_{t_1}^{t_2} \omega(t)\, dt\, \right |,
\label{rate0}
\end{equation}
where $t_1$ and $t_2$ are the dominant (closest to the real axis) turning points of the corresponding scattering potential. In the low frequency field with symmetrical chirp, more complex conjugate pairs appear with increasing chirp as demonstrated for various chirp values in Fig.~\ref{turning point} for $p_{\perp}=0$.

The multiple turning point generalizations of the pair production rate Eq.~\eqref{rate0} is
\begin{equation}
N_{\rm}\approx \sum_{i=1}^ne^{-2K_i}\pm \sum_{i\neq j}^n2\cos(2\alpha)\,e^{-K_i-K_j},
\label{rate}
\end{equation}
where $+$ and $-$ signs represent the scalar and spinor cases respectively. The other quantities are defined as
\begin{eqnarray}
\alpha&=& L-\sum_{i=1}^n\sigma(K_i)\nonumber, \\
L &=&\left| {\mathcal Re}\left( \int_{t_i}^{t_j} \omega(t)\, dt\right) \right | (i\neq j)\nonumber,\\
\sigma(K)&=&\frac{1}{2}\left[\frac{K}{\pi}\left(\ln \left(\frac{K}{\pi}\right)-1\right)+{\rm Arg}\,\Gamma\left(\frac{1}{2}-i\frac{K}{\pi}\right)\right]\nonumber.
\end{eqnarray}
The second term in Eq.~\eqref{rate} is the interference term and is responsible for the oscillations observed on the momentum spectrum.
The turning point pair numbers are increased with increasing chirp (see Fig.~\ref{turning point}) so that one can observe stronger interference effect on momentum spectrum for large chirp in Fig.~\ref{omg01pi05}(c) and (d). On the other hand, the symmetry of turning point is destroyed by increasing chirp, which results in an asymmetry more and more on the momentum spectrum in Fig.~\ref{omg01pi05}(c) and (d).
In addition, the dominant contribution to pair production rate comes from the terms involving turning points which are closest to the real axis and may cause enhancement. In Fig.~\ref{turning point}, we can also observe that the central pair of turning points (central pair on the right side) are more closest to real axis with increasing chirp. Therefore the corresponding pair production rate in Eq.~\eqref{rate} is always enhanced with the increasing chirp.

There exists various (competing) pair creation mechanisms in our numerical results corresponding to different chirp values. In low frequency field without chirping, pair production process is dominated by the tunneling mechanism. With increasing chirp, the field frequency contains many high-frequency components which are higher than its own original center frequency. In this way, the symmetrical chirp of low frequency field have played a role of the dynamically assisted Sauter-Schwinger mechanism such that it can produce more particles. In high frequency field, however, multiphoton absorption overtakes the process and the effect of dynamically assisted mechanism would become very weak. From this point of view, the chirp effect may be more helpful for pair production in the low frequency field.

\section{Summary}\label{Summary}

We have numerically studied the symmetrical chirp effects on the momentum spectrum and total particle number in two oscillating forms (cosine and sine) with spatial inhomogeneity. We have examined impacts of different field parameters, such as field central frequency and its chirp, spatial scale, and oscillating form on pair creation. Our results are summarized briefly as follows.

For high frequency field, the interference effect in the momentum spectrum is strengthen as chirp increases, and the total particle number is also enhanced greatly with increasing chirp. Also, the enhancement due to large chirp is more pronounced at smaller spatial scales and the momentum spectrum is more sensitive to the chirp in the cosine oscillation compared to the sinusoidal one.

For low frequency field, tunneling dominates the particle creation process and chirp plays crucial role. Small chirp has stronger interference effect on the momentum spectrum at larger spatial scales, while large chirp always causes obvious interference effects regardless of the spatial scale. It is noticed that the particle number increases much as chirp becomes large and, at small spatial scale, is enhanced by two orders of magnitude for large chirp.

More importantly, by comparing with the result of the usual asymmetrical frequency chirp, we have found that the total particle number is about two times larger in the symmetrical chirp case with the exception of the low frequency cosine oscillation form where it is almost the same.

These results indicate that the chirp symmetry and carrier phase play very important role on pair production in spatially inhomogeneous electric fields. While we have only considered two typical cases of the carrier phase, we speculate that more detailed analysis on the phase effect with symmetrical chirp may reveal further novel aspects about Schwinger pair creation.

\begin{acknowledgments}
\noindent
We thank M. Ababekri for critical reading of the manuscript. We are also grateful to Li Wang and N. Abdukerim for fruitful discussions and their help with the numerical calculation. This work was supported by the National Natural Science Foundation of China (NSFC) under
Grant No.\ 11875007 and No.\, 11935008. The computation was carried out at the HSCC of the Beijing Normal University.
\end{acknowledgments}

\end{document}